\documentclass[11pt]{article}
\usepackage{amsmath, amssymb, fontenc,  physics, mathtools}
\usepackage[superscript]{cite}
\usepackage{moreverb,url, multicol, multirow, soul,  lscape, longtable,booktabs,threeparttable, makecell, bm}
\usepackage[colorlinks,bookmarksopen,bookmarksnumbered,citecolor=blue,urlcolor=blue]{hyperref}
\usepackage{filecontents}
\usepackage{graphicx, caption, subcaption}
\usepackage{xcolor}
\usepackage{authblk}
\usepackage[top=1in, bottom=1in, left=1.25in, right=1.25in]{geometry}
\begin{document}
	
	\title{Regression with a right-censored predictor, \\using inverse probability  weighting methods.%\protect\thanks{This is an example for title footnote.}
			}
	
%	\author[1,2]{Roland A. Matsouaka*}
\author{
	Roland A. Matsouaka\footnote{corresponding author: 	Roland A. Matsouaka ({\color{blue} roland.matsouaka@duke.edu})}\\ Department of Biostatistics and Bioinformatics\\
	\& Program for Comparative Effectiveness Methodology, Duke Clinical Research Institute\\Duke University, Durham, North Carolina, USA\\
%	\\\href{mailto:me@somewhere.com}
	Email: {\color{darkgray} roland.matsouaka@duke.edu} 
	\and
	Folefac D. Atem\\Department of Biostatistics and Data Science \\University of Texas Health Science Center at Houston, Houston, Texas, USA \\%\href{mailto:me@somewhere.com}{me@somewhere.com} 
	Email: {\color{darkgray} folefac.atem@utsouthwestern.edu} 
}	

\maketitle	
	\begin{abstract}{In a longitudinal study, measures of  key variables might be incomplete or partially recorded due to drop-out, loss to follow-up, or early termination of the study occurring before the advent of the event of interest. In this paper, we focus primarily \textbf{on} the  implementation  of a regression model with a randomly censored predictor. We examine, particularly,  the 
		use of inverse probability weighting methods in a  generalized linear  model (GLM), when the predictor of interest is right-censored, to adjust for censoring. To improve the performance of the complete-case analysis and prevent selection bias, we consider three different weighting schemes: inverse censoring probability weights, Kaplan-Meier weights, and Cox proportional hazards weights. We use Monte Carlo simulation studies to evaluate and compare the empirical properties of different weighting estimation methods. Finally, we apply these methods to the Framingham Heart Study data as an illustrative example to estimate the relationship between age of onset of a clinically diagnosed cardiovascular event and low-density lipoprotein (LDL) among cigarette smokers.}
\end{abstract}	
%	\jnlcitation{\cname{%
%			\author{Williams K.}, 
%			\author{B. Hoskins}, 
%			\author{R. Lee}, 
%			\author{G. Masato}, and 
%			\author{T. Woollings}} (\cyear{2016}), 
%		\ctitle{A regime analysis of Atlantic winter jet variability applied to evaluate HadGEM3-GC2}, \cjournal{Q.J.R. Meteorol. Soc.}, \cvol{2017;00:1--6}.}

%	\footnotetext{\textbf{Abbreviations:} ANA, anti-nuclear antibodies; APC, antigen-presenting cells; IRF, interferon regulatory factor}
	%%%%%%%%%%%%%%%%%%%%%%%%%%%%%%%%%%%%%%%%%%%% %%%%%%%%%%%%%%%%%%%%%%%%%%%%%%%%%%%%%%%%%%%%%%%%%%%%%%%%%%%%%%%%%% %%%%%%%%%%%%%%%%%%%%%%
	%%%%%%%%%%%%%%%%%%%%%%%%%%%%%%%%%%%%%%%%%%%% %%%%%%%%%%%%%%%%%%%%%%%%%%%%%%%%%%%%%%%%%%%%%%%%%%%%%%%%%%%%%%%%%% %%%%%%%%%%%%%%%%%%%%%%
	%%%%%%%%%%%  Intro
	%%%%%%%%%%%%%%%%%%%%%%%%%%%%%%%%%%%%%%%%%%%% %%%%%%%%%%%%%%%%%%%%%%%%%%%%%%%%%%%%%%%%%%%%%%%%%%%%%%%%%%%%%%%%%% %%%%%%%%%%%%%%%%%%%%%%
	%%%%%%%%%%%%%%%%%%%%%%%%%%%%%%%%%%%%%%%%%%%% %%%%%%%%%%%%%%%%%%%%%%%%%%%%%%%%%%%%%%%%%%%%%%%%%%%%%%%%%%%%%%%%%% %%%%%%%%%%%%%%%%%%%%%%
	\section{Introduction}\label{sec:intro}
	Incomplete data, whether missing or censored, are ubiquitous in virtually all scientific disciplines, especially in economic, epidemiological, biological, biomedical, and environmental research studies. While several previously proven methods have been developed in the study of missing data and time to event outcomes, most of the literature on censored covariates deals the issue of the limit of detection. \cite{atem2017linear, kong2016semiparametric,atem2019cox} Nevertheless, over the past 5 years, Atem, Betensky, and colleagues have published a series of papers on randomly censored covariates. \cite{atem2017linear,atem2016multiple, qian2018threshold, atem2019improved}
	
	As opposed to a variable censored due to the limit of detection---where censoring is the result of inadequate instrument sensitivity to capture and quantify appropriately assay measures below (or beyond) some detectable limit\cite{schisterman2010opening}--- randomly censored variable measurements arise when there is a time lag or limited follow up  between the time when the variable is measured  and the occurrence of an event of particular interest that needs to happen for such a measurement to be accessible.\cite{atem2017linear,atem2019improved} For instance, when investigating familial aggregation in chronic disease incidence, Clayton  modeled  the possible influence  parental age at the onset of a given disease  might have on an individual's risk of succumbing to a particular disease.\cite{clayton1978model}  Atem and colleagues  also considered the relationship between the maternal age of onset of dementia, particularly at a younger age, and amyloid deposition burden in offspring over the age of 60 years.\cite{atem2017linear}
	%Using the Framingham Heart Study---an ongoing multi-generational landmark  study designed to identify factors and characteristics that contribute to the development of cardiovascular disease (CVD) and other diseases through long-term, active surveillance and monitoring---Atem and Matsouaka \cite{atem2016improving} studied the impact that age at onset of clinically-diagnosed cardiovascular events in parents may have on the  onset of cardiovascular events among offspring. 
	
	Interestingly, issues about or pertaining to data analysis with a potentially randomly censored predictor are not limited to studies involving  more than one generation (parent and offspring) of participants.  In clinical research, for instance, a censored predictor may be present if one desires to estimate the effects of timely reperfusion via percutaneous coronary intervention (PCI) on hospital outcomes for patients presenting a ST-elevation myocardial infarction. Measures of time to PCI can be censored for several reasons, including the absence of PCI-capable hospitals in the area, door-to-balloon time greater than 90 minutes, and patient arriving at the off-hours. \cite{hira2016temporal}  Censored predictors might also arise when investigating the association between the age of onset of cardiovascular disease (CVD) and the age of first cigarette smoking, the onset of chronic obstructive pulmonary disease (COPD) and the number of years of smoking,  or the relationship between the age of  CVD onset and low-density lipoprotein (LDL) among cigarette smokers, as we will present in this paper. For the latter example, while risk factors and history of CVD can be ascertained, it is unlikely that all the participants would have developed CVD at the time of the investigation. Thus,  the variable "age of onset of CVD" is not always guaranteed to be measured for all participants, i.e., it usually not be fully observed.  
	
	Common analysis methods for handling censored covariates include the complete-case analysis, simple substitution methods (also known as ad hoc "fill-in" methods), and threshold methods. The complete case analysis---which consists of discarding censored observations and running the analyses on a subset of the data without censored observations---is generally easy to use and may sometimes provide unbiased results. However, it reduces the sample size, wastes valuable information by discarding observations,  and often leads to inefficient results, especially under moderate or heavy censoring.\cite{atem2017linear,atem2019cox} Simple substitution methods---where censored observations are replaced by a constant, the overall mean, or the median of the covariate---has the advantage of using the entire dataset, but makes the implicit assumption that these substituted values are actual (known) values as if they were never censored. However, treating substituted values as known may result in a biased  and inefficient estimators with overly narrow confidence intervals since the methods tend to underestimate the true variability of the estimators. As argued by Helsel,  there is no theoretical justification for such methods; their use is akin to data fabrication and thus should be eschewed.\cite{helsel2006fabricating} Threshold methods consist either of dichotomizing the randomly censored covariate\cite{atem2017linear, austin2004estimating, rigobon2009bias} or using a specific threshold (when available) to extract information relative to the covariate via dichotomization.\cite{qian2018threshold} While dichotomizing continuous covariates is already a bad idea,\cite{royston2006dichotomizing, irwin2003negative, fitzsimons2008death,maccallum2002practice} it does not help that much when the variable is censored.\cite{austin2004estimating,rigobon2009bias} Finally, it is worth mentioning that, in the case of a time-to-event predictor  and a linear dependent variable, reversing the independent and dependent variable will result in a correct hypothesis test but but not in an unbiased parameter  estimates.\cite{atem2017linear}
	
	More appropriate methods for randomly censored covariates have been inspired by methods for missing data. Indeed, censored covariates  can be viewed as missing data with a specific structure since censored measures are known to lie beyond certain limiting values, even though the precise measures  are unknown. Therefore, methods for randomly censored covariates include maximum likelihood estimation, \cite{atem2017linear, tsimikas2012inference} estimating function approach,\cite{tsimikas2012inference} single and multiple imputations, as well as reverse survival regression methods.\cite{ atem2017linear, atem2016multiple,atem2019improved} Under some specific conditions, imputation methods consist in filling censored values with some plausible values  that represent (or replace) validly the censored observations. The uncertainty inherent to the imputation is judiciously incorporated in the analysis process to produce estimates and confidence intervals that reflect such an uncertainty. 
	
	Aside from the aforementioned methods, another framework that has been leveraged  in the literature to handle missing data is weighting. Weighting allows us to correct for selection bias that may have been introduced by using complete-case analysis, i.e., if we restrict the analysis solely to completely observed data. Similarly, even when the regression model is correctly specified, the presence of a censored predictor or of censored predictors can lead to biased and inefficient estimates, especially if the percentage of censored observations is high  or the censoring is informative, i.e., depends on patients' demographic and medical history.  Therefore, weighting improves our analysis by assigning a weight to each participant based on the occurrence of the event of interest which is used to improve efficiency via a weighted generalized linear model (GLM).  
	
	The intuitive idea of a weighting method is to define and estimate sampling weights based on the censoring mechanism to change the composition of sub-sample of patients with non-censored covariate measures. It generates a pseudo-population in which  all non-censored participants represent themselves and provide additional copies of themselves, where the number of copies reflects the magnitude of their corresponding weight.\cite{hernan2019causal}  Running such a weighted GLM may help reduce bias due to censoring and improve the precision of the complete-case analysis. Clearly, as in the context of missing data, using the weighting methods  can also be a better alternative to  censored covariate imputation.\cite{seaman2013review}  
	
	However, to our knowledge,  there is a paucity of publications that address the issue of randomly censored covariates using weighting methods. The only exceptions are the papers on treatment-duration policy for which a dynamic-treatment-regimes perspective was used by Johnson and Tsiatis, \cite{johnson2004estimating} along with a handful of applications and extensions.\cite{george2016cancer}  Their questions of interest were "what is the optimal dose or  how long treatment must be administered to see meaningful results?" On one hand, if stopped early, patients will experience minor  or no side effects at all, the effect of the drug will not manifest and the drug will be wasted. On the other hand, administering a high dose of the drug may be more effective in treating the condition, but more and more patients might experience  serious side effects and complications, eventually leading to treatment discontinuation. The goal was to find the optimal treatment infusion duration that can be recommended to inform future guidelines and clinical policy.
	
	In this paper, we consider inverse probability weighting methods to estimate parameters of a GLM with a randomly censored covariate.  These methods restrict their attention to observations with complete covariate measures and weight each of their contributions to obtain consistent estimators. The goal is to up weight observed data to compensate and account for the loss of information due to censored observations. 
	For this purpose, we discuss three different weighting methods and consider two types of censoring; one being the random censoring (or administrative censoring), which occurs at the end of the study  and the other, the informative censoring, where the censoring mechanism depends on  patients' (measured) characteristics and outcomes. We start in Section \ref{sec:ipw} with some notation and provide an overview of the weighting methods in GLM. Then, we introduce the use of the inverse probability censoring weighting (IPCW) where the weights are defined via the predicted probabilities of being uncensored. Next, leveraging the structure of the censored covariate in Section \ref{sec:kaplan}, we derive non-parametric weights using the Kaplan-Meier (IPCW KM) estimates of censoring times. Furthermore, assuming proportional hazards of censoring times among patients, we develop a method to determine Cox proportional hazards related weights (IPCW Cox) in Section \ref{sec:cox}. Furthermore, we evaluate and compare the three methods, along with the complete-case (CC) analysis, through simulation studies in Section \ref{sec:simul}, looking at both independent and dependent censoring. Finally, in Section \ref{sec:apply}, we apply the four methods (CC, IPCW, IPCW KM, and IPCW Cox) to assess the relationship between age of parental onset of a clinically diagnosed cardiovascular event and low-density lipoprotein in offspring amongst smokers using data from the Framingham Heart Study.\cite{mahmood2014framingham,tsao2015cohort} In section \ref{sec:discussion}, we conclude the paper with some discussion points .
	
	%%%%%%%%%%%%%%%%%%%%%%%%%%%%%%%%%%%%%%%%%%%%%%%%%%%%%%%%%%%%%%%%%%%%%%%%%%%%%%%%%%%%%%%%%%%%%%%%%%%%%%%%%%%%%%%%%%%%%%%%%%%%%%%%%%%%%%
	%%%%%%%%%%%%%%%%%%%%%%%%%%%%%%%%%%%%%%%%%%%%%%%%%%%%%%%%%%%%%%%%%%%%%%%%%%%%%%%%%%%%%%%%%%%%%%%%%%%%%%%%%%%%%%%%%%%%%%%%%%%%%%%%%%%%%%
	%%%%%%%%%%%%%%%%%%%%%%%%%%%%%%%%%%%% Inverse probability censoring weights (IPCW) %%%%%%%%%%%%%%%%%%%%%%%%%%%%%%%%%%%%%%%%%%%%%%%%%%%%
	%%%%%%%%%%%%%%%%%%%%%%%%%%%%%%%%%%%%%%%%%%%%%%%%%%%%%%%%%%%%%%%%%%%%%%%%%%%%%%%%%%%%%%%%%%%%%%%%%%%%%%%%%%%%%%%%%%%%%%%%%%%%%%%%%%%%%%
	%%%%%%%%%%%%%%%%%%%%%%%%%%%%%%%%%%%%%%%%%%%%%%%%%%%%%%%%%%%%%%%%%%%%%%%%%%%%%%%%%%%%%%%%%%%%%%%%%%%%%%%%%%%%%%%%%%%%%%%%%%%%%%%%%%%%%% 
	\section{Inverse probability of censoring weights (IPCW)}\label{sec:ipw}
	\subsection{Notation, definition, and identification}\label{sec:note}
	We consider a generalized linear model specified by 
	\begin{align}\label{eq:glm}
	&E(Y_i|X_i=x_i, \mathbf{Z}_i=\mathbf{z}_i)=\mu_i,  ~ i=1,\ldots, n \nonumber\\
	%E(\mathbf{Y}|\mathbf{X}X_i=x_i, z_i)&=\mu_i 
	&g(\mu_i)=\eta_i=\beta_0+\beta_1x_i+\beta_2'\mathbf{z}_i,  ~\text{with}~ \mathbf{z}_i=(z_{i1}, \dots, z_{ip})\nonumber\\
	&Var(Y_i|X_i=x_i,  \mathbf{Z}_i=\mathbf{z}_i)=\tau^2v(\mu_i). 
	\end{align}
	The variable $Y$ is the outcome of interest, $g(.)$ is the link function,  $X$ is the continuous variable   that might be randomly censored for some participants, $\mathbf{Z}$ is  the matrix of fully measured covariates of dimension $n\times p$, $\beta=(\beta_0, \beta_1,  \beta_2')'$ is the  $(p+2)$-length column vector of regression model coefficients to estimate,   $v(\mu)$ is the variance function, and $\tau^2$ is the dispersion parameter. 
	
	The choice of the link function $g(.)$ depends on the nature of the outcome of interest. For the linear regression with a continuous outcome,  the identity function $g(u)=u$, while $g(u)=\log(u)$ and $g(u)=log(u/(1-u))$ correspond, respectively, to the log link for count data outcome and logit link for binary outcome. We say that the model \eqref{eq:glm} is correctly specified if  $\beta$ exits such that the equation is satisfied for all values of $X$ and $\mathbf{Z}$.   
	
	For each subject $i$ in the observed data consist of $O_i=(V_i, \Delta_i, \mathbf{Z}_i, Y_i),$ where $V_i=min(X_i, C_i)$, with $C_i$ being the censoring variable and $\Delta_i$ the censoring indicator, i.e. , $\Delta_i=1$ if the $i$-th subject's covariate value is observed and $\Delta_i=0$ if the value is censored. We assume that both the censoring time $C$ and censored covariate $X$ are continuous.
	
	In the absence of censoring, the generalized estimation equation for ${\beta}$ is 
	\begin{equation}
	%\mathbf{U(\beta)}=\sum\limits_{i=1}^{n}h(X_i\mathbf{Z}_i; \beta)\{y_i-g^{-1}(\beta_0+\beta_1x_i+\beta_2'\mathbf{z}_i)\}X
	\mathbf{U(\beta)}=\sum\limits_{i=1}^{n}U_i(\beta)=\sum\limits_{i=1}^{n}h(X_i,\mathbf{Z}_i; \beta)\{(y_i-\mu_i)\}\mathcal{X}_i=0 \label{eq:nocens}
	\end{equation}
	where $h(X_i,\mathbf{Z}_i; \beta)={\tau^2v(\mu_i)}^{-1}\frac{\partial }{\partial\beta} [g^{-1}(\eta_i)]$ is a given scalar function and $\mathcal{X}_i=(1,x_i,\mathbf{z}'_i)'$.  The function $h$ is equal to ${\tau^2v(\mu_i)}^{-1}$ for the identity link and $1$ for both the log and the logit link functions. Other functions $h$ are presented and discussed by McCullagh and Nelder\cite{mccullagh1989generalized} as well as by Liang and Zeger;\cite{liang1986longitudinal}  Depending on the nature of the outcome $Y$, we can derive a closed-form solution for ${\beta}$ (e.g., with the identity link) or use an iterative algorithm  to solve the estimating equation (e.g., using the logit link) \eqref{eq:nocens}.
	 
	Since $E(\mathbf{U(\beta)})=0$, by standard asymptotic arguments,\cite{mccullagh1989generalized, liang1986longitudinal} the solution to the estimating equation \eqref{eq:nocens}, denoted by $\widehat{\beta}$, is an $\sqrt{n}$-consistent estimator of the true parameter of interest $\beta$ i.e. 
	$\displaystyle 
	\sqrt{n}(\widehat{\beta}-\beta)\longrightarrow MVN(0, A_0^{-1}B_0[A_0^{-1}]'),
	$
	where the matrices $A_0$ and $B_0$, respectively of size $(p+2)\times n$ and $n\times n$, are given by
	\begin{align*}
	A_0=-E\left[\frac{\partial U_i}{\partial \beta'}\right]_{\widehat{\beta}=\beta}~\text{and}~B_0=E\left[U_i(\beta)U_i'(\beta)\right]_{\widehat{\beta}=\beta}.
	\end{align*}
	and $A_0^{-1}B_0[A_0^{-1}]'$ is the asymptotic variance-covariance  matrix for $\widehat{\beta}$ of size $(p+2)\times (p+2).$
	
	\noindent
	In the presence of censoring, i.e., when not all $x_i$ are observed,  $$\displaystyle \mathbf{U}(\beta)=\sum\limits_{i=1}^{n}[\Delta_iU_i(\beta)+(1-\Delta_i)U_i(\beta)].$$ Therefore, we can no longer solve appropriately the  modified estimation equation \eqref{eq:nocens}. 
	The complete-case analysis is based on the data $\left\{O_i=(V_i, \Delta_i, \mathbf{Z}_i, Y_i)
	: \Delta=1, i=1, \ldots, n\right\}$ where we ignore (remove the data) observations with censored values. Although the complete-case analysis is easy to implement, it results in a loss of efficiency due to data deletion. Also, it may yield inconsistent parameter estimates and lead to spurious results, especially when the censoring mechanism is covariate-dependent. \cite{hernan2004structural} 
	
	To circumvent such limitations, we restrict our attention to subjects with a fully observed $X,$ but their contribution is leveraged using the selection probability $\pi(Y_i, \mathbf{H}_i; \theta)=P(\Delta=1|Y_i, \mathbf{H}_i; \theta)$ as a weight to correct for a possible selection bias created by censoring, where $\theta$ is the associated selection parameter and $\mathbf{H}$ contains $\mathbf{Z}$ as a subset. The vector $\mathbf{H}$ may include, in addition to $\mathbf{Z}$, auxiliary variables that are associated  with $X,$ $Y,$ and the censored indicator $\Delta$ in such a way that $C$ and $X$ are independent given the covariate vector $\mathbf{H}$  and the response $Y.$ By auxiliary variables, we refer to variables that do not provide additional information to the regression model \eqref{eq:glm}, but are considered potentially informative about and predictive of the censoring $C$, which can help reduce bias due to censoring.\cite{robins1992recovery,robins1994estimation} The choice of variables to include in the selection model $\pi(Y_i, \mathbf{H}_i; \theta)$ are beyond the scope of this paper. We suppose throughout this paper that there is a set of fully observed auxiliary variables. Interested readers can refer to papers that have extensively investigated  the issue and proposed necessary steps to make such a choice,\cite{schafer1999multiple,collins2001comparison, schafer2003multiple, yuan2014consistency, raykov2016enhancing, enders2016multiple} which  closely resemble the steps required to select variables to include in propensity score models to adjust for confounding or selection bias.\cite{brookhart2006variable, greenland2015statistical}
	
	The idea of using weights based on the selection probability is reminiscent of the Horwitz and Thompson's principles of sampling design in the analysis of survey data.\cite{horvitz1952generalization} This idea has regained attention in statistical literature of missing and censored  data in recent years.\cite{koul1981regression, robins1992recovery,robins1994estimation,seaman2013review, bartlett2014improving} If $\pi(Y_i, \mathbf{H}_i; \theta)$ is known, the estimating equation for $\beta$ is given by
	\begin{equation}
	%\mathbf{U(\beta)}=\sum\limits_{i=1}^{n}h(X_i\mathbf{Z}_i; \beta)\{y_i-g^{-1}(\beta_0+\beta_1x_i+\beta_2'\mathbf{z}_i)\}X
	0=\sum\limits_{i=1}^{n}U_i^{icpw}(\beta)=\sum\limits_{i=1}^{n}W_i\Delta_i U_i(\beta), ~~\text{where }~~ W_i=\pi(Y_i, \mathbf{H}_i; \theta)^{-1}.\label{eq:cens}
	\end{equation}
	As long as $\pi(Y_i, \mathbf{H}_i; \theta)>0,$ 
	\allowdisplaybreaks\begin{align*}
	\displaystyle \displaystyle E\left[\sum\limits_{i=1}^{n}W_i\Delta_i  U_i(\beta)\right]&=E\left[\sum\limits_{i=1}^{n}\frac{\Delta_i}{\pi(Y_i, \mathbf{H}_i; \theta)} U_i(\beta)\right]\\&=E\left[E\left[\sum\limits_{i=1}^{n}\frac{\Delta_i}{\pi(Y_i, \mathbf{H}_i; \theta)} U_i(\beta)\right]|Y_i, \mathbf{H}_i\right]\\
	&=E\left[\sum\limits_{i=1}^{n}\frac{E\left[\Delta_i|Y_i, \mathbf{H}_i\right]}{\pi(Y_i, \mathbf{H}_i; \theta)} U_i(\beta)\right]=E\left[\sum\limits_{i=1}^{n} U_i(\beta)\right]=0.
	\end{align*}
	Therefore, the solution to the estimating equation \eqref{eq:cens} is a consistent and asymptotically normal estimator of $\beta$. In practice, however, $\pi(Y_i, \mathbf{H}_i; \theta)$ is unknown in most cases and must be estimated.  In the next sections, we will determine the solution to the generalized estimate \eqref{eq:cens} given a specification of the selection probability model $\pi(Y_i, \mathbf{H}_i; \theta).$
	
	%%%%%%%%%%%%%%%%%%%%%%%%%%%%%%%%%%%%%%%%%%%% %%%%%%%%%%%%%%%%%%%%%%%%%%%%%%%%%%%%%%%%%%%%%%%%%%%%%%%%%%%%%%%%%% %%%%%%%%%%%%%%%%%%%%%%
		\subsection{IPCW via a logistic regression model}\label{sec:lr}
	%%%%%%%%%%%%%%%%%%%%%%%%%%%%%%%%%%%%%%%%%%%% %%%%%%%%%%%%%%%%%%%%%%%%%%%%%%%%%%%%%%%%%%%%%%%%%%%%%%%%%%%%%%%%%% %%%%%%%%%%%%%%%%%%%%%%
	One of the methods to estimate the selection probability $\pi(Y_i, \mathbf{H}_i; \theta)$ is via logistic regression. In that case, an estimate of $\pi(Y_i, \mathbf{H}_i; \theta)$ is given by 
	
	\allowdisplaybreaks\begin{align*}
	\widehat{\pi}\left(Y_i, \mathbf{H}_i; \widehat{\theta}\right)=P(\Delta_i=1|Y_i, \mathbf{H}_i)=\frac{\exp(\widehat{\theta}_0+\widehat{\theta}_1Y_i+\widehat{\mathbf{\theta}}_2'\mathbf{H}_i)}{1+\exp(\widehat{\theta}_0+\widehat{\theta}_1Y_i+\widehat{\mathbf{\theta}}_2'\mathbf{H}_i)}
	\end{align*}
	where $\widehat{\theta}=(\widehat{\theta}_0, \widehat{\theta}_1,\widehat{\theta}_2)$ is the estimator of the parameter $\theta$ from the logistic regression model.
	
	The selection probability $\widehat{\pi}\left(Y_i, \mathbf{H}_i; \widehat{\theta}\right)$ thus is defined as the probability of observing a measured (i.e., non-censored) $X$ given the outcome $Y$ and the covariate vector $\mathbf{H}$. Since the logistic regression model is based on the binary censoring indicator $\Delta$, this weighting scheme resembles that of the propensity score inverse probability weighting method. The basic intuition is that weighting any subject in the complete-case sample by the weights $\displaystyle\widehat{W}_i=\widehat{\pi}\left(Y_i, \mathbf{H}_i; \widehat{\theta}\right)^{-1}$ is equivalent to  allowing such a subject to represent him- or herself  and $\displaystyle\left(\widehat{W}_i-1\right)$ other subjects who have the same propensity (or more generally have similar characteristics) to be censored as those for whom the variable $X$ is censored and thus not observed. Hence, in the causal inference parlance, weighting via  the selection probability  $\widehat{\pi}\left(Y_i, \mathbf{H}_i; \widehat{\theta}\right)$ creates a pseudo-population (i.e., the weighted sample) which consists of copies of subjects whose measures of the covariate $X$ are not censored. %Robins and colleagues have shown that using stabilized weights have important efficiency advantages while still providing consistent parameter estimators.\cite{robins1992recovery, robins1994estimation}
	
	%%%%%%%%%%%%%%%%%%%%%%%%%%%%%%%%%%%%%%%%%%%% %%%%%%%%%%%%%%%%%%%%%%%%%%%%%%%%%%%%%%%%%%%%%%%%%%%%%%%%%%%%%%%%%%%%%%%%%%%%%%%%%%%%%%%%
	\subsection{IPCW via the Kaplan Meier Estimator}\label{sec:kaplan}	
	%%%%%%%%%%%%%%%%%%%%%%%%%%%%%%%%%%%%%%%%%%%% %%%%%%%%%%%%%%%%%%%%%%%%%%%%%%%%%%%%%%%%%%%%%%%%%%%%%%%%%%%%%%%%%%%%%%%%%%%%%%%%%%%%%%%%
	Unlike the IPCW above, which is based on logistic regression and only uses limited information as to whether an observation is censored or not, our second approach is based on the (nonparametric) Kaplan Meier estimator, which is particularly used in the analysis of time-to-event data with censored observations.  %a very robust and proven technique to predict the propensity of certain event, this approach was based on
	
	 The corresponding selection probability $\pi(Y_i, \mathbf{H}_i; \theta)$ can be estimated as $\widehat{\pi}\left(Y_i, \mathbf{H}_i; \theta\right)=\widehat{K}(X_i)$ where $\widehat{K}(X_i)$ is the Kaplan-Meier estimator for the survival function of the censoring time and $K(u)=P(C>u),$  the cumulative probability of remaining uncensored from the time  $u$ onward. Note that the Kaplan-Meier estimator $\widehat K(.)$ is based on the data $(V_i, \Delta^*)$ where $\Delta^*=1-\Delta$, which means $\widehat K(.)$ is  determined by reversing the roles of survival times $X_i$ and censoring times $C_i$, i.e. the survival time $T_i$ censors the censored time $C_i$.\cite{zhao1997consistent, lin2000linear, anstrom2001utilizing,  bang2002median}
	
	This estimation method of the selection probability via the Kaplan-Meier estimator is unbiased and yields reliable results when the censoring mechanism is purely random, i.e., $C$ is independent of $X$, $\mathbf{H}$, and the response $Y.$ The potential time to censoring is defined using the survivor function $K(u)=P(C>u).$ Such a censoring occurs when values of $X$ are potentially censored only through administrative censoring, which is strictly related to the study design.  %This Kaplan-Meier (KM) estimator, a robust and proven technique to predict a (time) distance to an event, without taking into consideration other coveriates. Therefore, for each subject with observed value $X_i$, the probability of not being censored is  $K(X_i)$. 
	
	%%%%%%%%%%%%%%%%%%%%%%%%%%%%%%%%%%%%%%%%%%%% %%%%%%%%%%%%%%%%%%%%%%%%%%%%%%%%%%%%%%%%%%%%%%%%%%%%%%%%%%%%%%%%%%%%%%%%%%%%%%%%%%%%%%%%	
	\subsection{IPCW using Cox proportional hazards model} \label{sec:cox}
	%%%%%%%%%%%%%%%%%%%%%%%%%%%%%%%%%%%%%%%%%%%% %%%%%%%%%%%%%%%%%%%%%%%%%%%%%%%%%%%%%%%%%%%%%%%%%%%%%%%%%%%%%%%%%%%%%%%%%%%%%%%%%%%%%%%%
	Valid inference from a Kaplan-Meier estimator might be possible only under the assumption that no additional covariate is related to the censoring and the censoring mechanism is independent. In other words, the distribution of censored observations is similar to the distribution of uncensored observations. If the censoring mechanism is informative, i.e., when censoring is strongly related to some other covariate(s), we need to control for or associated with the outcome since the Kaplan-Meier estimator may no longer be unbiased and optimal, which may lead to biased estimates of the selection probability. In that case, the parameter of the GLM model estimated using the above Kaplan-Meier estimator of the selection probability might be biased since such an estimator may not be sufficient enough to predict censoring adequately.  
	
	Similar to the KM estimator, the Cox proportional hazards model (Cox model) approach is another technique to predict the time to an event, under the proportional hazards assumption. It incorporates into its prediction model valuable information that is provided by all the measured potential confounders. This method works when the censoring mechanism is independent or when the censoring mechanism $C$ may depend on the set of covariates $\mathbf{H}$ and even indirectly on $Y$.\cite{robins1992recovery,robins1993information,robins2000correcting} The latter property is extremely valuable, especially when the measured covariates are strongly connected with censoring and thus might contain additional information about the probability of censoring.
	
	Using the time-to-event framework for our censored variable, we can use a Cox proportional hazards model the same way it is commonly used in survival analysis, to adjust for informative censoring---along with substantial loss of information when there is a high percentage of censoring during the study follow-up. In that case, assuming that the censoring times are continuous, the selection probability $\pi(Y_i, \mathbf{H}_i; \theta)$ can be estimated by the way of the corresponding hazard function related to the censoring events 
	\allowdisplaybreaks\begin{align*}
	\lambda_C(u|\mathbf{H}_i; \theta)=\lambda_0(u)\exp({\theta}_1Y_i+{\mathbf{\theta}}_2'\mathbf{H}_i),
	\end{align*}
	where $\lambda_0(u)$ is the baseline hazard function. 
	In that case, the selection probability is determined by the cumulative conditional probability that an individual will be uncensored through time $u$  given  $Y_i$ and $\mathbf{H}_i$, %which can be estimated by the following extension of the Kaplan-Meier product limit estimator for censoring. 
	\allowdisplaybreaks\begin{align*}
	&\pi(u|Y_i, \mathbf{H}_i; \widehat\theta)=P(C>u|Y_i=y_i,\mathbf{H}_i=\mathbf{h}_i; \widehat\theta)\\
	&= \prod_{s<u}\left[1-d\Lambda_C\left(u|Y_i=y_i,\mathbf{H}_i=\mathbf{h}_i; \widehat\theta\right)\right]=\!\!\!\prod_{\{j;V_j<u, \Delta_j=0\}}\left[1-\widehat{\lambda}_0(V_j; \widehat\theta)\exp\left\{\widehat{\theta}_1Y_i+\widehat{\mathbf{\theta}}_2'\mathbf{H}_i\right\}\right]
	\end{align*}
	 where  $\widehat{\theta}=(\widehat{\theta}_1,\widehat{\mathbf{\theta}}_2)$ is an estimator of  ${\theta}=({\theta}_1,{\mathbf{\theta}}_2)$, $\displaystyle\Lambda_C(u|Y_i=y_i,\mathbf{H}_i=\mathbf{h}_i; \widehat\theta)=\int_{0}^{u}\lambda_C(s|\mathbf{H}_i; \widehat\theta)ds$ is the cumulative hazard, and $\displaystyle\widehat{\lambda}_0(V_j; \widehat\theta)=\displaystyle{\Delta^*}\left[{\displaystyle\sum_{k=1}^{n}I(V_k\ge V_j)\exp\left(\widehat{\theta}_1Y_k+\widehat{\mathbf{\theta}}_2'\mathbf{H}_k\right)}\right]^{-1}$ is the Breslow  estimator of the baseline hazard function $\lambda_0$.\cite{breslow1972contribution}
	
	%and note that K(t) is essentially the probability that the ith person survived to time t without being censored (Robins and Rotnitzky, 1992).

	Similar to the weights from logistic regression, the weights defined herein allows subjects with an uncensored measure of $X$ who resemble (with respect to $Y$ and $\mathbf{H}$) those with censored measures  to receive more weights. Therefore, at each time $u$, a subject with an observed measure of $X$ can be considered as representing  $W(u)=\pi(Y_i, \mathbf{H}_i; \widehat\theta)^{-1}$ individuals in the pseudo population, including him- or herself.\cite{robins1993information,robins2000correcting} 
	\subsubsection{Remarks}
	\begin{enumerate}
		\item Even when the selection probabilities $\pi(Y_i, \mathbf{H}_i; \widehat\theta)$  are known, it has been shown that we gain efficiency by using estimated probabilities instead of the true probabilities.\cite{robins1994estimation, robins1995analysis} A heuristic explanation of this seemingly counterintuitive result is that estimating the selection probabilities enables one to use all the available data more efficiently. Indeed, such an estimation incorporates more effectively measured covariates information on both censored and uncensored observations along with auxiliary variables, significant interactions, and high-order polynomial terms as well as the outcome  $Y$.  
		
		\item In two of the examples we mentioned in the introduction, it is reasonable to assume that the censoring mechanism to which a covariate $X$ measured on parents is unrelated to the outcome  $Y$ measured on their offspring. However, such an independence assumption cannot be generalized in all situations and should be considered only based on expert opinion, prior knowledge, informed understanding of the design, the predictor(s) of interest, and the outcome under study. 		
		Such an ascertainment of the reasons for censoring was conducted, for example, by the investigators of the ESPRIT (Enhanced Suppression of Platelet IIb/IIIa Receptor with Integrillin Therapy) study. In ESPRIT, the objective was to evaluate the effect of the duration of a continuous infusion of Integrellin for 18--24 hours in patients on a composite endpoint of mortality, myocardial infarction, and target urgent vessel revascularization.\cite{johnson2004estimating} For patients who experienced serious complications during treatment, the infusion process was discontinued to provide appropriate medical attention. Therefore, when measures of the duration of the treatment were censored, it was at random, independently of the expected patient’s outcome(s).
		
		In general, given that the censoring time $C$ and the variable $X$ are independent given $\mathbf{Z}$, we can show that 
		\allowdisplaybreaks\begin{align}\label{eq:y}
		f(Y|V,\Delta=1, \mathbf{Z})&=\frac{f(Y, X=v, C>v |\mathbf{Z})}{f(X=v, C>v|\mathbf{Z})} %\nonumber\\&
		=\frac{f( X=v, C>v |Y,\mathbf{Z})f(Y,|\mathbf{Z})}{f(X=v, C>v|\mathbf{Z})} \nonumber\\
		&=\frac{f( X=v|Y,\mathbf{Z})P(C>v |Y,\mathbf{Z})f(Y|\mathbf{Z})}{f( X=v|\mathbf{Z})P(C>v|\mathbf{Z})} \nonumber\\&
		=\frac{f( X=v|Y,\mathbf{Z})f(Y|\mathbf{Z})}{f( X=v|\mathbf{Z})}\frac{Pf(C>v |Y,\mathbf{Z})}{P(C>v|\mathbf{Z})} \nonumber\\
		&=f( Y|X=v,\mathbf{Z})\frac{S_C(v |Y,\mathbf{Z})}{S_C(v|\mathbf{Z})}, ~\text{with}~ S_C(v|.)=P(C>v |.)
		\end{align}
		%\begin{align*}
		%f_{Y|V,\Delta=1, \mathbf{Z}}(y_i|X_i=x_i, \mathbf{Z}_i=\mathbf{z}_i)=\frac{f_Y(y_i|\mathbf{z}_i)}{f_{X|\mathbf{Z}}(v_i|\mathbf{z}_i)f(C>v_i|\mathbf{z}_i)}\int_{v_i}^{\infty}f_{X, C|Y, \mathbf{Z}}(t_i, c|y_i,\mathbf{z}_i)
		%\end{align*}
		provided that $C$ and  $X$ are also independent given $Y$ and $\mathbf{Z}$.
		
		Equation \eqref{eq:y} provides a theoretical justification of using the outcome $Y$ in the selection model when such dependence is warranted directly or indirectly.  Such a modeling strategy, far from being a self-fulfilling prophecy, has become ubiquitous in the imputation procedures of missing data to ensure congeniality between the imputation and analysis models.\cite{moons2006using,little1992regression,harel2007multiple, schafer2003multiple, meng1994multiple} 
		
		%In addition, if we assume that the censoring time $C$ and the variable $X$ are independent given $Y$ and $\mathbf{Z},$
		%\begin{align*}
		%.f_{Y|V,\Delta=1, \mathbf{Z}}(y_i|X_i=x_i,\Delta=1, \mathbf{Z}_i=\mathbf{z}_i)&=\frac{f_Y(y_i|\mathbf{z}_i)}{f_{X|\mathbf{Z}}(v_i|\mathbf{z}_i)P(C>v_i|\mathbf{z}_i)}\int_{v_i}^{\infty}f_{X, C|Y, \mathbf{Z}}(v_i, c|y_i,\mathbf{z}_i)dc\\
		%&=\frac{f_{X|Y, \mathbf{Z}}(v_i|y_i,\mathbf{z}_i)f_Y(y_i|\mathbf{z}_i)}{f_{X|\mathbf{Z}}(v_i|\mathbf{z}_i)P(C>v_i|\mathbf{z}_i)}\int_{v_i}^{\infty}f_{C|Y, \mathbf{Z}}(c|y_i,\mathbf{z}_i)dc\\
		%&=\frac{f_{Y|X, \mathbf{Z}}(y_i|v_i,\mathbf{z}_i)}{P(C>v_i|\mathbf{z}_i)}S_{C|Y, \mathbf{Z}}(c|y_i,\mathbf{z}_i)dc
		%\end{align*}
		%\item {\color{blue} On the use of auxiliary variables to improve the estimation of selection probabilities} The use of the outcome $Y$ in the selection probability model can be mitigated by the use of auxiliary covariates. These are variables which are not part of regression model 
		
		\item {\bf On stabilized weights:} Even when the condition that $\pi(Y_i, \mathbf{H}_i; \widehat\theta)>0$ holds,  the parameter estimator $\beta$ can be very unstable and perform poorly under small and moderate sample sizes if, for some subjects, the selection probability $\pi(Y_i, \mathbf{H}_i; \widehat\theta)$ is very small.\cite{robins2000marginal, hernan2006estimating, schafer2008average} To mitigate the influential effects of large weights, Robins and Hernan recommend using {\it stabilized weights}, which consist of replacing the selection weights $W_i$ by $SW_i=f(C)\times W_i$ where $f(C)$ is the expected value of being uncensored.\cite{hernan2019causal} If we considered weights from the logistic regression, $f(C)=P(\Delta=1).$   For the IPCW weights estimated via the Cox proportional hazards model, the expected value $f(C)$  is replaced by $f(C|u)$ which is derived using the usual Kaplan-Mieir estimator $f(C|u)=\widehat{K}^0(u)=P(C>u)$ (i.e., the probability of  being uncensored from the time $u$ onward), \cite{robins1993information, robins2000marginal, hernan2006estimating, robins2000correcting} as indicated in Section \ref{sec:kaplan}.	
		\item {\bf On Cox proportional hazards model:} Although Cox (semiparametric) proportional   hazards model is undoubtedly the most popular method of analysis for right-censored data, there are other alternative parametric and semiparametric methods that can be used when appropriate (see Andersen and Keiding\cite{andersen1995survival} or Guo and Zheng,\cite{guo2014overview} along with the references therein).
		\item The above results can be extended easily to more multiple randomly censored covariates. Depending on the plausible assumptions we can make regarding the censoring distributions of these covariates, and estimate the selection probabilities  $\displaystyle\pi(Y_i, \mathbf{H}_i; \theta)=\prod_{k=1}^K\pi_k(Y_i, \mathbf{H}_i; \theta),$ where $K$ is the total number of randomly censored covariates. 
	\end{enumerate}
	%%%%%%%%%%%  Augmentation
	%\section{Improving efficiency through augmentation}\label{sec:aipw}
	%%%%%%%%%%%  Calculations
	%\section{Estimating the parameter coefficients in practice}\label{sec:estim}
	%%%%%%%%%%%  Simulations
	\section{Simulation study}\label{sec:simul}
	We conducted simulation studies to investigate the performance of each of the proposed IPW methods. For comparison, we also included both the full model (Full), i.e. the model without censoring and the complete-case (CC) method, to highlight the extent to which discarding observations impacts the results. 
	Firstly, we generated two sets of data corresponding to two types of censoring (independent censoring or outcome-dependent censoring). For each set of data, we considered two different degrees of censoring, labeled "light" and "heavy" censoring, with  20\% and 40\% of censored observations, respectively. 
	
	\subsection{Simulation setup}
	In each dataset,  we  generated two fully observed covariates $Z_1$ and $Z_2$, with $Z_1\sim N(18.5,3)$ and $Z_2\sim Ber(0.5)$. We then generated the censored covariate  $X\sim \textit{Weibull}  (0.2, 0.25)$, and the outcome $Y=\beta_{0}+\beta_{1}Z_{1}+\beta_{2}Z_{2}+\beta_{3}X+\varepsilon$,  with $\varepsilon\sim N(0, 0.1)$. We set $\beta_{0}=0.005$, $\beta_{1}=0.01$, $\beta_{2}=-0.01$, and $\beta_{3}=-0.05$.  Finally, we considered different scenarios for the distribution of the censoring mechanism $C$ of $X$ as indicated in Table \ref{tab:censoring}. For each scenario of the censoring mechanisms, we generated $M=5,000$ data sets of size  $n=400$ and $n=600$, respectively.
		\begin{table*}[h] %\small
		\caption{Distribution of the censoring variable $C$}\label{tab:censoring}
		\begin{center} 
			\begin{threeparttable}
					\def\arraystretch{1.2}
				\begin{tabular}{ccccccccc} 
					\toprule
				&     	 \multicolumn{2}{@{}c@{}}{\textbf{Censoring type}}  \\\cmidrule(lr){2-3}
					%&   &  	 \multicolumn{4}{@{}c@{}}{\textbf{N = 400}} & & \multicolumn{4}{@{}c@{}}{\textbf{N = 600}} \\\cmidrule(lr){3-6} \cmidrule(lr){8-11}
					\textbf{}	     & 	\multicolumn{1}{@{}c@{}}{\textbf{Independent censoring}}   &  \multicolumn{1}{@{}c@{}}{\textbf{Outcome-dependent censoring}} \\\midrule 
				Full	&   \multirow{1}{*}{No censoring} &   \multirow{1}{*}{No censoring} \\\addlinespace%\cmidrule(lr){2-3}
					\multirow{2}{*}{Light}  &  \multirow{2}{*}{$C\sim \textit{Weibull}  (1, 2)$}  &  $C_{1}\sim \textit{Weibull}  (1, 2)$; $~C_{3}\sim \textit{Weibull}  (1.5, 2);$\\%\addlinespace 
				(20\% censored)	&  &    $C= I(\varepsilon> 0)C_1+ I(\varepsilon\leq 0)C_3$ \\ \addlinespace %\cmidrule(lr){2-3}
%					&   &   & &   &  \\%\addlinespace
					\multirow{2}{*}{Heavy}  & \multirow{2}{*}{$C\sim \textit{Weibull}  (1, 0.35)$}   & $C_{2}\sim \textit{Weibull}  (1, 0.35)$;  $~C_{4}\sim \textit{Weibull}  (1.5, 0.35);$ \\%\addlinespace 
					(40\% censored)&  &    $C= I(\varepsilon> 0)C_2+ I(\varepsilon\leq 0)C_4$ \\ %\addlinespace
						\bottomrule
				\end{tabular} 
				
				\begin{tablenotes}
					\footnotesize
					\item Note: $I(.)$ is the standard indicator function
				%	\item CC: complete case; IPCW: inverse probability of censoring weights; KM: Kaplan Meier.
				\end{tablenotes}
			\end{threeparttable}
		\end{center} 	
	\end{table*} 	
\subsection{Evaluation criteria} 
 From each generated data set, we run a weighted regression model---based on the aforementioned weighting methods---and  determined the estimates $\widehat \beta_{3m}$, $m=1,\ldots, M,$  of the parameters of interest $ \beta_{3}$   and considered the overall estimate $\overline{\widehat \beta}_3=\displaystyle{M}^{-1}\!\sum_{m=1}^M{\widehat \beta_{3m}}$. Then, we calculated   five different measures of performance: the  $\displaystyle\text{bias}=(\overline{\widehat \beta}_{3}- \beta_{3})$; the  bias percentage $\%Bias=100|{ \text{Bias}}/{\beta_{3}}|$; the model-based standard error    (SE) (i.e., the average of all standard errors of $\beta_{3}$ from each fitted model); the Monte Carlo simulation standard deviation (SD) (i.e.,  the empirical standard error of the estimates of $\beta_{3}$ over all $M$ simulated data sets);    and  the mean squared error (MSE = Bias$^2 + $ SE$^2$). \cite{morris2019using} The most consistent and efficient method will be the one with the smallest bias and MSE as well as similar SD and SE. Naturally, we expected the scenario with no censoring (full data) to yield results with the most consistent and efficient model parameter. Therefore, the performance of the different methods were then assessed based on how close the were from the full model. 

%{\color{blue} \underline{Note to Roland} If the notation SE and SD are reversed as I am proposing here, then I will have to update the tables too.}
\subsection{Simulation results}\label{simulation_results}
Tables \ref{tab1:wgts_summary} and \ref{tab2a:wgts_summary} summarize the results from our simulation study. They indicate, as expected, that the performance of the different methods (along with the CC analysis) depends on  the sample size, the proportion of censored observations on the potential censored predictor, and the type of of censoring mechanism. For each method, at a given  percentage of the censored predictor, the bias as well as the percentage decreases in  magnitude as the sample size increases. Also, for a given sample size, the bias increases as the percentage of censored covariate increases from $20\%$ to $40\%$. 

For Table \ref{tab1:wgts_summary}, the censoring mechanism is independent, i.e., it is neither a function of other covariate(s) nor the dependent variable. These simulated results for sample sizes $n=400$ and $n=600$, showed that the three IPCW methods together with the CC approach were marginally  biased. The CC approach and the IPCW Cox are least  biased.  In the former, censored observations are deleted that might result to drop in power. Thus, the standard error of the CC analyses were high and increased tremendously as the censoring rate increased. This explains the high MSEs for the complete-case analysis when censoring is heavy as compared to the other the three IPCW methods. The IPCW Cox performed best as compared to the other IPCW approaches because it used all available information to determine its weights. The IPCW based on logistic regression to predict the propensity of $X$ not being censored  and the IPW based on Kaplan Meier to predict the distance of an event without taking into consideration other covariates appeared to be  indistinguishable. These approaches resulted in the highest bias and MSE. The IPCW based on Cox model that tries to construct the weight slightly differently by predicting a (time) distance to an event while taken into consideration all the variables, instead of the propensity of event performed the best. Its bias and MSE are the smallest. 

Thus, in the case of independent censoring, even though censoring is completely random, the (final) outcome $Y$ still depends indirectly on this random process and only the IPCW Cox takes into consideration all these components. In other words, the IPCW Cox uses all the information at hand while the other approaches use partial information. The IPCW Cox takes explicitly into consideration the time at censoring, all possible available predictors and the outcome $Y$, which is included as a predictor in the Cox model, to come up with an optimal weights. Although the IPCW adjust also for possible available predictors and the outcome, the logistic regression model used to estimate the weights is based on a binary indicator of whether censoring occurs or not. This logistic model does not adjust for the time at which each censored event occurs. Whether censoring happens at the very beginning of the follow-up period or not, it is treated exactly the same as the one occurring towards the end of the follow-up period. Thus, the loss of information and  precision inherent to the IPCW method.  Finally, while the IPCW KM accounts for when the censoring event occurs, it does not factor in the information provided by  the available predictors or the outcome, leading also to less precise estimates of the censoring weights compared to the IPCW Cox. In the case of missing data, the IPCW approach uses all available information and should produce optimal weights. Since we do not have a time-to-event scenario in missing data, the IPCW Cox should not be considered in such a context. 

For Table \ref{tab2a:wgts_summary}, when censoring is dependent on the outcome, that is, the censoring mechanism is a function of the dependent variable. These simulated results for sample sizes $n=400$ and $n=600$ showed significant differences between the three ICPW methods together with the CC approach in terms of the bias and MSE. The bias of the CC, IPCW, and IPCW KM increases tremendously as censoring increased from light to heavy. The MSE of these three approaches also experienced a similar increase. The performance of the IPCW Cox is slightly biased as compared to that of the Full model. The bias and MSE reduce with an increased sample size from $n=400$ to $n=600$, but increased slightly as censoring increased from light to heavy. It is also worth mentioning that, as for heavy censoring, the model-based standard errors (SE) and the simulation standard deviation (SD) for the proposed IPCWs are not very close for the heavy censoring cases. This is due to a reduction in precision as censoring increases. Thus, the Kaplan-Meier estimator and Cox proportional hazard model suffered a reduction in power in estimating its parameters, thus reducing the precision of the weights.
	%\subsubsection{Independent censoring}\label{sec:indpt}
%	\begin{center}
%		\begin{table*}[t]%
%			\caption{Simulation results.*\label{tab1}}
%			\centering
%			\begin{tabular*}{500pt}{@{\extracolsep\fill}lccD{.}{.}{3}c@{\extracolsep\fill}}
%				\toprule
%				&\multicolumn{2}{@{}c@{}}{\textbf{Spanned heading\tnote{1}}} & \multicolumn{2}{@{}c@{}}{\textbf{Spanned heading\tnote{2}}} \\\cmidrule{2-3}\cmidrule{4-5}
%				\textbf{col1 head} & \textbf{col2 head}  & \textbf{col3 head}  & \multicolumn{1}{@{}l@{}}{\textbf{col4 head}}  & \textbf{col5 head}   \\
%				\midrule
%				col1 text & col2 text  & col3 text  & 12.34  & col5 text\tnote{1}   \\
%				col1 text & col2 text  & col3 text  & 1.62  & col5 text\tnote{2}   \\
%				col1 text & col2 text  & col3 text  & 51.809  & col5 text   \\
%				\bottomrule
%			\end{tabular*}
%			\begin{tablenotes}%%[341pt]
%				\item Source: Example for table source text.
%				\item[1] Example for a first table footnote.
%				\item[2] Example for a second table footnote.
%			\end{tablenotes}
%		\end{table*}
%	\end{center}

	\begin{table*}[h] %\small
		\caption{Simulation results for independent censoring*}\label{tab1:wgts_summary}
			\begin{center} 
			\begin{threeparttable}
				\begin{tabular}{ccrrccrrcccccccccccc} 
					\toprule
					&   &  	 \multicolumn{4}{@{}c@{}}{\textbf{$n = 400$}} &  \multicolumn{4}{@{}c@{}}{\textbf{$n = 600$}} \\\cmidrule(lr){3-6} \cmidrule(lr){7-10}
					%&   &  	 \multicolumn{4}{@{}c@{}}{\textbf{N = 400}} & & \multicolumn{4}{@{}c@{}}{\textbf{N = 600}} \\\cmidrule(lr){3-6} \cmidrule(lr){8-11}
					 Censoring  & \textbf{Method} &	 Bias (\%) & SE & SD & MSE        &   Bias( \%)  & SE & SD & MSE     \\\midrule 
					&  Full  &  0.01 (2)&  0.48 & 0.48   & 0.23  &  0  (0) & 0.38  & 0.39 &0.15  \\%\addlinespace
					\multirow{3}{*}{Light}  & CC &  0.01  (2)  & 1.20& 1.21 & 0.15  & 0.01  (2)& 0.98  & 0.98 &0.97  \\%\addlinespace 
					  &  IPCW  &       -0.10  (20)&      0.41 & 0.21 & 1.17  &  -0.10 (20) & 0.38 & 0.19 & 1.16\\ %\addlinespace
					&   IPCW KM    &  0.09    (18)  &  0.52& 0.38 & 1.10  &  0.08  (16) &0.41 &0.23 & 0.81\\ %\addlinespace
					&   IPCW Cox    &  0.02    (4)&     0.49& 0.59 & 0.28  &  0.01 (2) & 0.39&0.47  & 0.16\\ \addlinespace \cmidrule(lr){2-10}
					&  Full &  0.01 (2)& 0.48 & 0.48    &  0.23 &     0  (0) &0.38 &  0.39   &  0.15   \\%\addlinespace
					\multirow{3}{*}{Heavy}  & CC &  0.07  (14)  & 3.23& 3.21 & 11  &  0.04  (8) & 2.63& 2.64  &  7.08  \\%\addlinespace 
					  &  IPCW  &       0.18    (36) &    1.48 &  2.63 & 5.44  &     0.18   (36)& 1.09  &2.18 &  4.43 \\ %\addlinespace
					&   IPCW KM    &  -0.19    (38)  &  1.73& 2.41  & 6.60  &  0.18  (36) & 1.13  &2.21 &   4.52 \\ %\addlinespace
					&   IPCW Cox    &  0.03    (6) &    1.42& 2.17 & 2.11  &    0.03 (6) &      1.32& 2.11 & 1.83 \\ 
					\bottomrule
				\end{tabular} 
		
				\begin{tablenotes}
					\footnotesize
					\item  *{\bf Note}: Bias, SD, and SE provided in $10^{-1}$; MSE provided in $10^{-4}$;  
					\item  CC: complete-case analysis; IPCW: inverse probability of censoring weights; 
					 IPCW KM: IPCW via Kaplan Meier estimator; IPCW Cox: IPCW via a Cox proportional hazards model for censoring.
				\end{tablenotes}
			\end{threeparttable}
	    	\end{center} 	
	\end{table*} 

	%\subsubsection{Comments}

	%\subsection{Covariate-dependent censoring}\label{sec:dpt}
	\begin{table}[h] %\small
		\caption{Simulation results for outcome dependent censoring*} \label{tab2a:wgts_summary}
		\begin{center} 
			\begin{threeparttable}[]
				\begin{tabular}{ccrccccrccccccccccccccccccccccccc} 
					%\multicolumn{1}{c}{\bf PART}  &\multicolumn{1}{c}{\bf DESCRIPTION} \\ 
					\toprule
					&   &  	 \multicolumn{4}{@{}c@{}}{\textbf{$n = 400$}} &  \multicolumn{4}{@{}c@{}}{\textbf{$n = 600$}} \\\cmidrule(lr){3-6} \cmidrule(lr){7-10}
					%&   &  	 \multicolumn{4}{@{}c@{}}{\textbf{N = 400}} & & \multicolumn{4}{@{}c@{}}{\textbf{N = 600}} \\\cmidrule(lr){3-6} \cmidrule(lr){8-11}
					 Censoring    & \textbf{Method} &	 Bias (\%)   &SE & SD & MSE         &  Bias (\%) & SE & SD & MSE     \\\midrule 
					&  Full & -0.01 (2)& 0.48 & 0.48   & 0.24  &   0.01 (2) & 0.38  & 0.39 &0.15  \\%\addlinespace
					\multirow{3}{*}{Light}  & CC &  -0.06 (12) &   1.26 & 1.24 & 1.90  &0.02 (4) & 1.01  &  1.01  & 1.10  \\%\addlinespace 
					  &  IPCW  &       0.20    (40)  &   0.49 & 2.01 & 4.20  & 0.20  (40)& 0.39 & 1.78 & 4.20\\ %\addlinespace
					&   IPCW KM    & -0.23   (46) &   0.57& 2.51 & 5.60  &  -0.21 (42) &0.45 &2.25 & 4.60\\ %\addlinespace
					&   IPCW Cox    &  -0.04 (8)&      0.48& 0.71 & 0.39  &  0.04  (8)&0.38&0.61  & 0.30\\\addlinespace \cmidrule(lr){2-10}
					&  Full &  -0.01  (2)&  0.48 & 0.48    &  0.24 &    0.01 (2) &0.38   &0.39   &  0.15   \\%\addlinespace
					\multirow{3}{*}{Heavy}  & CC &  0.23  (46)   & 3.57& 3.51 & 18.0  &  0.13  (25) & 2.88& 2.91 &  9.90  \\%\addlinespace 
					  &  IPCW  &       -0.33    (66) &  1.66 &  2.35 & 14.0 &    -0.26  (52)& 1.16  &1.18 &  8.11 \\ %\addlinespace
					&   IPCW KM    &  -0.30    (59) &   1.65& 4.44  & 12.0  &  -0.29  (58) & 1.33  &4.16 &   10.3 \\ %\addlinespace
					&   IPCW Cox    &  0.09   (18) &    1.46& 2.01 & 2.90  &     0.08   (16) &  1.16& 1.84 & 2.00 \\ 
					\bottomrule
				\end{tabular} 
				\begin{tablenotes}
					\footnotesize
					\item *{\bf Note}: Bias, SD, and SE provided in $10^{-1}$; MSE provided in $10^{-4}$. 
					\item  CC: complete-case analysis; IPCW: inverse probability of censoring weights; 
					  IPCW KM: IPCW via Kaplan Meier estimator; IPCW Cox: IPCW via a Cox proportional hazards model.
				\end{tablenotes}
			\end{threeparttable}
		\end{center} 
	\end{table}

    \subsection{Additional simulations }
    \subsubsection{Simulations to mimic the real data example }
    Similar to the real data example, we generated two fully observed covariates $Z_1$ and $Z_2$, with $Z_1\sim Ber(0.53)$ and $Z_2\sim Ber(0.52)$. We then generated the potential censored covariate  $X\sim \textit{Uniform}  (0.3, 1.30)$, and the outcome $Y=\beta_{0}+\beta_{1}Z_{1}+\beta_{2}Z_{2}+\beta_{3}X+\varepsilon$,  with $\varepsilon\sim N(0, 0.01)$. To mimic the real data analysis,  we increased the sample size to $n=850$ and chose the variance of $\varepsilon$ to be 0.01, that is, the variance of the outcome is about $10$ times smaller than in the previous two simulated. We set $\beta_{0}=4.90$, $\beta_{1}=0.0037$, $\beta_{2}=0.10$, and $\beta_{3}=0.045$.  Finally, we considered different scenarios for the distribution of the censoring mechanism $C$ of $X$, we set $C=C_{1}\sim \textit{Weibull} (0.75, q)$ if $Z_1=0$ and $C=C_{2}\sim \textit{Weibull} (1.25, q)$ if  $Z_1=1$, where $q=2.50$, $1.50$, and $0.70$ for $20\%$, $40\%$, and $65\%$ censoring respectively. For each covariate dependent censoring mechanisms, we generated $M=5,000$ data sets of sample size $n=850$. We also simulated data with an interaction term between the potential censored covariate $X$ and $Z_1$. The regression coefficient  $(\beta_{4})$ for this interaction term was set at $0.05$. All other parameters were kept as described in the linear regression without an interaction.
    
   \subsubsection{Simulation results to mimic the real data example }
     
   Table \ref{tab3:wgts_summary} summarizes the results from this covariate dependent simulated data. The choice of the variance of the outcome together with the increase in the sample size of the simulated data resulted to increase precision. The four approaches, that is, CC, IPCW, IPCW KM, and IPCW Cox result in relatively unbiased estimates for the three proportions of censoring. Nonetheless, as censoring increases to $65\%$ the three IPCWs approaches percentage bias increase significantly from less than $10\%$ to $13.5\le bias \le 18.9$, with the least percentage bias from the IPCW Cox and the highest from the IPCW. However, based on MSE, the effect of these biased estimates from the IPCW KM and IPCW Cox approaches are minimal, since estimates from these approaches perform better than minimal percentage bias CC approach. The results from the model with the interaction term are presented in Table \ref{tab2:wgts_summary} in  the Appendix. 
   
 		\begin{table}[h] %\small
 		\caption{Simulation results: A covariate dependent censoring to mimic the real-data example} \label{tab3:wgts_summary}
 		\begin{center} 
 			\begin{threeparttable}[]
 				\begin{tabular}{rcrccccrccccccccccccc} 
 					%\multicolumn{1}{c}{\bf PART}  &\multicolumn{1}{c}{\bf DESCRIPTION} \\ 
 					\toprule
 					&   &  	 \multicolumn{5}{@{}c@{}}{\textbf{$n = 850$}} & %& \multicolumn{5}{@{}c@{}}{\textbf{$n = 600$}} 
 					\\\cmidrule(lr){3-7} %\cmidrule(lr){9-13}
 					%&   &  	 \multicolumn{4}{@{}c@{}}{\textbf{N = 400}} & & \multicolumn{4}{@{}c@{}}{\textbf{N = 600}} \\\cmidrule(lr){3-6} \cmidrule(lr){8-11}
 					\textbf{Censoring}	     & \textbf{Method} &	 Bias  & 	\% Bias  &SE & SD & MSE        %& &  Bias & 	\% Bias & SE & SD & MSE     
 					\\\midrule 
 					&  Full & 0	 & 0 & 	0.12 & 	0.12 & 	1.4 %  &  &0.01 &2 & 0.38  & 0.39 &0.15  
 					\\%\addlinespace
 					\multirow{3}{*}{20\%}  & CC & -0.001 & 	0.28 & 	0.14 & 	0.14 & 	4.1 % & &0.02 &4 & 1.01  &  1.01  & 1.10  
 					\\%\addlinespace 
 					&  IPCW  &  0.02 & 	5.4 & 	0.14 & 	0.31 & 	2.0 %& & 0.20 & 40& 0.39 & 1.78 & 4.20
 					\\ %\addlinespace
 					&   IPCW KM    & 0.004 & 	0.99 & 	0.14 & 	0.14 & 	2.0 % % & & -0.21 &42 &0.45 &2.25 & 4.60
 					\\ %\addlinespace
 					&   IPCW Cox    &  0.004 & 	1.09 & 	0.14 & 0.14 & 	0.2 %
 					 % & & 0.04 & 8&0.38&0.61  & 0.30
 					\\\addlinespace \cmidrule(lr){2-7}
 					&  Full & 0	 & 0 & 	0.12 & 	0.12 & 	1.4 	%  &  &0.01 &2 & 0.38  & 0.39 &0.15  
 					\\%\addlinespace
 					\multirow{3}{*}{40\%}  & CC &  0.02 & 	5.4	 & 0.16 & 	0.16 & 	2.6 % & &0.02 &4 & 1.01  &  1.01  & 1.10  
 					\\%\addlinespace 
 					&  IPCW  &       0.03 & 	8.1	 & 0.16 & 	0.33 & 	2.7 %& & 0.20 & 40& 0.39 & 1.78 & 4.20
 					\\ %\addlinespace
 					&   IPCW KM    & -0.02 & 	5.4 & 	0.15 & 	0.17 & 	2.3 % & & -0.21 &42 &0.45 &2.25 & 4.60
 					\\ %\addlinespace
 					&   IPCW Cox    &  -0.02 & 	5.2 & 	0.15 & 	0.16 & 	2.4 % & & 0.04 & 8&0.38&0.61  & 0.30
 					\\\addlinespace \cmidrule(lr){2-7}
 					&  Full &  0 & 	0 & 	0.12 & 	0.12 & 	1.4 %  &   &  0.01 &2 &0.38   &0.39   &  0.15 
 					\\%\addlinespace
 					\multirow{3}{*}{65\%}  & CC &  0.04 & 	10.8 & 	0.21 & 	0.21 & 	4.6 %& & 0.13 & 25 & 2.88& 2.91 &  9.90  
 					\\%\addlinespace 
 					&  IPCW  &       0.07 & 	18.9 & 	0.21 & 	0.23 & 	4.9 % &   &  -0.26  & 52& 1.16  &1.18 &  8.11 
 					\\ %\addlinespace
 					&   IPCW KM    &  0.06 & 	16.2 & 	0.16 & 	0.21 & 	2.9 % & & -0.29 & 58 & 1.33  &4.16 &   10.3 
 					\\ %\addlinespace
 					&   IPCW Cox    &  0.05	 & 13.5	 & 0.16 & 	0.21 & 	2.8 % & &    0.08  & 16 &  1.16& 1.84 & 2.00 
 					\\ 
 					\bottomrule
 				\end{tabular} 
 				\begin{tablenotes}
 					\footnotesize
 					\item *{\bf Note}: Bias, SD, and SE provided in $10^{-2}$; MSE provided in $10^{-6}$. 
 					CC: complete-case analysis; IPCW: inverse probability of censoring weights; IPCW KM: IPCW via Kaplan Meier estimator; 
 					 IPCW Cox: IPCW via a Cox proportional hazards model.
 				\end{tablenotes}
 			\end{threeparttable}
 		\end{center} 
 	\end{table} 
	
%	\subsection{Comments and remarks}
	
	%%%%%%%%%%%%%%%%%%%%%%%%%%%%%%%%%%%%%%%%%%%% %%%%%%%%%%%%%%%%%%%%%%%%%%%%%%%%%%%%%%%%%%%%%%%%%%%%%%%%%%%%%%%%%% %%%%%%%%%%%%%%%%%%%%%%
	%%%%%%%%%%%%%%%%%%%%%%%%%%%%%%%%%%%%%%%%%%%% %%%%%%%%%%%%%%%%%%%%%%%%%%%%%%%%%%%%%%%%%%%%%%%%%%%%%%%%%%%%%%%%%% %%%%%%%%%%%%%%%%%%%%%%
	%%%%%%%%%%%  Application
	%%%%%%%%%%%%%%%%%%%%%%%%%%%%%%%%%%%%%%%%%%%% %%%%%%%%%%%%%%%%%%%%%%%%%%%%%%%%%%%%%%%%%%%%%%%%%%%%%%%%%%%%%%%%%% %%%%%%%%%%%%%%%%%%%%%%
	%%%%%%%%%%%%%%%%%%%%%%%%%%%%%%%%%%%%%%%%%%%% %%%%%%%%%%%%%%%%%%%%%%%%%%%%%%%%%%%%%%%%%%%%%%%%%%%%%%%%%%%%%%%%%% %%%%%%%%%%%%%%%%%%%%%%
	\section{Application to the Framingham Health Study}\label{sec:apply}
	Cigarette smoking is a major risk factor for CVD.\cite{gossett2009smoking}  The adverse effects of smoking on CVD risk involve complex mediation through multiple interrelated mechanisms, including increased oxidative stress, endothelial injury and dysfunction, altered blood coagulation, altered metabolism, and derangements of lipid composition. \cite{gossett2009smoking}  The association between low-density lipoprotein (LDL) and CVD has been well studied and documented among individuals receiving certain therapeutic interventions, \cite{silverman2016association} but not much is known about this association among cigarette smokers.
	
	To assess this question, we compared the CC analysis to the three different IPCW methods using data from the Framingham Heart Study (FHS) database, a well-known premier, longitudinal prospective cohort study  for studying cardiovascular diseases (CVD) and other terminal diseases.\cite{mahmood2014framingham,tsao2015cohort} For this paper, we consider data from the Framingham Offspring Cohort (FHSO) launched in 1971 and whose participants have been examined, on average, every 3 to 4 years since enrollment. The FHSO dataset consists of a sample of 3,514 biological descendants of the Original FHS Cohort, 1,576 of their spouses and 34 adopted offspring for a total sample of 5,124 subjects ($48\%$ males). Our analyses used observations collected during the FHSO Exam 7 (1998--2001, $n = 3,539$) and FHSO Exam 8 (2007, $n=2,898$). After cleaning and merging the required data by including only smokers, the sample size was reduced to $n=886$, of which $292$ (32.96\%) were clinically diagnosed with CVD, $468$ (52.82\%) were regular beer consumers, and $466$  (52.60\%) were male.

	In Table \ref{data:apply1}, we provided a summary of all four approaches in the association between age of onset of CVD as a predictor and log(LDL) as a dependent variable, while controlling for sex (male versus female) and beer intake (yes or no) as confounders. As shown, the estimates of the age of onset of CVD differed across the different methods. The CC analysis estimate, based on about 33\% non-censored observations,  was greater than the IPCW but less IPCW KM and IPCW Cox.  For the IPCW approach, a logistic regression for observing a non-censored event was fitted with  beer intake, male, and LDL as covariates. For the IPCW KM the weight was derived through a   Kapla-Meier estimator approach for time to onset on CVD, while for the IPCW Cox, the weight was derived from the Cox regression model for time to onset of CVD with beer intake, male, and LDL as exploratory variables. 
	
	The estimates for beer intake and sex (male versus female) were similar for IPCW KM and IPCW Cox but not for CC or IPCW. Interestingly, the parameter of interest, age of onset of CVD, was significant for IPCW KM and IPCW Cox ($p<0.0001$) but not for CC or IPCW. The point  estimate for the ICPW KM and ICPW Cox method was significantly larger ($\approx 0.036$) than those of the other two methods (resp. $0.0014$ and $0.0002$ for CC and IPCW). 
	
	In the CC analysis method, deleting observations ($\approx 67\%$) resulted in a significant drop in precision as shown by the increased standard errors. As demonstrated by the simulation studies, the CC analysis method together with the IPCW and IPCW KM methods resulted in biased estimates when censoring depends on the outcome. However, when censoring is dependent on other covariates along with a sufficiently large number of events, as it is the case in this example, the IPCW KM results to estimates that are comparable to IPCW Cox. Since the censoring mechanism in the FHSO data likely depends on baseline variables, the IPCW KM and the IPCW Cox (unlike the CC and the IPCW) resulted in statistically significant estimates for the age of onset of CVD and thus provided a more realistic assessment of the relationship between the onset of CVD and LDL. Finally, the standard errors of the parameter of interest were roughly the same across all the three probability weighting methods.

\begin{table}[h]
	\caption{Relationship between age of onset of  CVD and LDL among cigarette smokers }\label{data:apply1}
	\begin{center}
		\begin{threeparttable}
			%\begin{tabular}{p{4cm}lcccc}
			\begin{tabular}{rrrrrrrrrrrrrrrrrrrr}
				\toprule
				Method & & Variable&  Estimate & SE & t-value  & p-value   \\ \midrule
				\multirow{4}{*}{CC*} %(32.96\% of the data)
				&	&Intercept & 5.0091 & 0.0535 & 93.58 & $< 0.0001$  \\
				&	&Age$^{\dagger}$ & 0.0014 & 0.0010 & 1.36& 0.1762  \\
				&	&Beer & 0.0195 & 0.0226 & 0.87 &0.3876  \\
				&	&Male & 0.0311 & 0.0234 & 1.33 &0.1855  \\\addlinespace
				
				\multirow{4}{*}{IPCW}& & Intercept & 5.0297 & 0.0328 & 153.19 & $< 0.0001$  \\
				&	&Age$^{\dagger}$ & 0.0002 & 0.0006 & 0.30& 0.7679  \\
				&	&Beer & 0.0158 & 0.0145 & 1.08 &0.2784  \\
				&	&Male & 0.0419 & 0.0145 & 2.89 &0.0040  \\\addlinespace
				
				\multirow{4}{*}{IPCW KM}& &Intercept & 4.8039 & 0.0198 & 242.45 & $< 0.0001$  \\
				&	&Age$^{\dagger}$ & 0.0036 & 0.0005 & 7.03&  $< 0.0001$  \\
				&	&Beer & 0.0531 & 0.0142 & 3.74 & 0.0002  \\
				&	&Male & 0.1079 & 0.0142 & 7.62 &$< 0.0001$ \\\addlinespace
				
				\multirow{4}{*}{IPCW Cox}& &Intercept & 4.8074 & 0.0203 & 237.12 & $< 0.0001$  \\
				&	&Age$^{\dagger}$ & 0.0037 & 0.0005 & 6.81 & $< 0.0001$ \\
				&	&Beer & 0.0431 & 0.0140 & 3.08 &0.0021 \\
				&	&Male & 0.1176 & 0.0138 & 8.50 &$< 0.0001$ \\\bottomrule
			\end{tabular}
			\begin{tablenotes}
				\footnotesize
				\item {\bf Note}: $^{\dagger}$Age at onset of CVD;  Beer = beer consumption; CC$^{*}$: complete-case analysis. 
				  IPCW: inverse probability of censoring weights; 
				IPCW KM: IPCW via Kaplan Meier estimator; IPCW Cox: IPCW via a Cox proportional hazards model.
			\end{tablenotes}
		\end{threeparttable}
	\end{center}
\end{table}
	
	%%%%%%%%%%%%%%%%%%%%%%%%%%%%%%%%%%%%%%%%%%%% %%%%%%%%%%%%%%%%%%%%%%%%%%%%%%%%%%%%%%%%%%%%%%%%%%%%%%%%%%%%%%%%%% %%%%%%%%%%%%%%%%%%%%%%
	%%%%%%%%%%%%%%%%%%%%%%%%%%%%%%%%%%%%%%%%%%%% %%%%%%%%%%%%%%%%%%%%%%%%%%%%%%%%%%%%%%%%%%%%%%%%%%%%%%%%%%%%%%%%%% %%%%%%%%%%%%%%%%%%%%%%
	%%%%%%%%%%%  Discussion
	%%%%%%%%%%%%%%%%%%%%%%%%%%%%%%%%%%%%%%%%%%%% %%%%%%%%%%%%%%%%%%%%%%%%%%%%%%%%%%%%%%%%%%%%%%%%%%%%%%%%%%%%%%%%%% %%%%%%%%%%%%%%%%%%%%%%
	%%%%%%%%%%%%%%%%%%%%%%%%%%%%%%%%%%%%%%%%%%%% %%%%%%%%%%%%%%%%%%%%%%%%%%%%%%%%%%%%%%%%%%%%%%%%%%%%%%%%%%%%%%%%%% %%%%%%%%%%%%%%%%%%%%%%
	\section{Discussion} \label{sec:discussion}
	%Standard regression methods used under the name of generalized linear models can typically lead to biased estimates whenever one of the predictors is randomly censored. As censoring may be strongly associated with other covariates and possibly correlated with the outcome, it can introduce selection bias.\cite{hernan2004structural} Thus, the  need to adjust for such a censoring to successfully recover adequate and robust parameter estimates.   
	
	In this paper, we have presented three different inverse probability weighting methods, together with analyses from the complete-case method to model data when the primary predictor of interest is right-censored. Throughout our simulation study and the analysis from the Framingham Heart Study data, the need to properly account for censoring and employing a statistical model that provides unbiased estimates of the parameter of interest irrespective of the type of censoring cannot be overemphasized. 
	 
	 The simulation study showed varying percentage bias for the four methods, i.e., complete-case analysis, inverse probability weight via a logistic regression model, the inverse probability weighting via Kaplan-Meier estimator and the inverse probability weighting via Cox model. For well-powered studies, all three IPCW approaches together with the CC resulted in less biased estimates when the censoring was either dependent on baseline covariates and completely independent. As precision dropped, the IPCW and IPCW KM resulted in biased estimates. Furthermore, when censoring was dependent on the outcome of interest, the CC, IPCW, and IPCW KM approach resulted in biased estimates. Since it is very difficult to have a sample where censoring is completely independent or dependent (of the other covariates and outcome), these approaches are likely to result in biased estimates and poor inferences.  
	 
	 This becomes more evident in the analysis of real-world data we considered. The estimates from CC and IPCW methods resulted in an insignificant association between the age of onset of CVD and LDL among cigarette smokers. These two approaches that resulted in biased estimates failed to take into consideration the time of censoring in the model. However, by accounting for both the variable dependence of the censoring mechanism and incorporating the actual time of censoring in its analytic approach, the inverse probability weight via KM and Cox model provided a significant estimate. Furthermore, it is worth mentioning that the inverse probability weight via the Kaplan-Meier estimator resulted in estimates that are like those based on the inverse probability weight via the Cox model. Nonetheless, since the inverse probability weight via Cox is the only reliable method that resulted in an unbiased estimate irrespective of the type of censoring mechanism, the estimates for the association between age of onset of CVD and LDL among smokers should be based on IPCW Cox.
	
	This work highlights the robustness provided by the distance (or time) to event inverse probability weighting method via the Cox model in estimating the effect of a potentially right-censored covariate on an outcome. Even though the weights estimated via a Kaplan-Meier estimator was also based on the distance to the event, this method does not consider possible measured confounders and, thus, lack of robustness when censoring is dependent on the outcome. Finally, the traditional inverse probability weight based on the propensity score of not being censored carries less information and appeared weaker when censoring is directly related to the outcome of interest.
	\bibliographystyle{unsrt}
	\bibliography{icpw_arxiv}
%	\bibliography{C:/Users/rm267/Dropbox/Work/Folefac/Censor_Paper//4thPaper/icpw_sim}
%	\begin{appendices}
%		\renewcommand\thetable{\thesection\arabic{table}}
%		\renewcommand\thefigure{\thesection\arabic{figure}}
%		%  \section{Consectetur adipiscing elit} \label{app:foobar}
		%    \begin{table}[h]
		%    \caption{foo}
		%    \begin{tabular}{cc}
		%    \textbf{ a } & \textbf{ b }\\
		%    1 & 3 \\
		%    2 & 4\\
		%    \end{tabular}
		%    \end{table}
		%  \section{Mauris euismod}
		
		%\renewcommand{\theequation}{A-\arabic{equation}}    
		%  % redefine the command that creates the equation no.        
		%  \section*{APPENDIX}  % use *-form to suppress numbering
		\setcounter{equation}{0}  % reset counter     
		\setcounter{section}{0} 
		\setcounter{table}{0} 
%		\numberwithin{equation}{section}
%		\numberwithin{table}{section}
		\noindent
		\section*{Appendix: Additional simulation results} \label{sec:app1} 
%		\\[2ex]
%		\subsection*{Appendix 1: Additional simulation results} 
	\begin{landscape}
	\begin{table}[ht] %\small
		\caption{Simulation results for a covariate dependent censoring with interaction between fully observed covariate (FOC) and a censored covariate ($n=850$)*} \label{tab2:wgts_summary}
		\begin{center} 
			\begin{threeparttable}[h]

				\begin{tabular}{ccrrccccrcccccrccrcccccccccccccc} 
					%\multicolumn{1}{c}{\bf PART}  &\multicolumn{1}{c}{\bf DESCRIPTION} \\ 
					\toprule
					&   &&  	 \multicolumn{4}{@{}c@{}}{\textbf{Potential censored covariate (PCC)}}  & & \multicolumn{4}{@{}c@{}}{\textbf{Interaction between FOC and PCC}} \\\cmidrule(lr){4-7} \cmidrule(lr){8-12}
					%&   &  	 \multicolumn{4}{@{}c@{}}{\textbf{N = 400}} & & \multicolumn{4}{@{}c@{}}{\textbf{N = 600}} \\\cmidrule(lr){3-6} \cmidrule(lr){8-11}
					\textbf{Censoring}	     & \textbf{Method}& &	 Bias	(\%)   &SE & SD & MSE       &   &  Bias  (\%)   & SE & SD & MSE     \\\midrule 
					&  Full && 	0  (0) &	$0.17$ &	$0.17$ &	$2.9$  &	& 0 (0)& 	0.24& 	0.24& 	5.8\\%\addlinespace
					\multirow{3}{*}{20\%}  & CC &&  0.01 (2.7) &$0.21$	 &$0.21 $&	$4.4 $ && 0.01  (0.2)	& 0.28& 	0.28& 	7.9 \\%\addlinespace 
					&  IPCW  &&       0.01  (2.7)	 &$0.24$ &	$0.27$ &$	5.8$  & &  0.08  (1.6)& 0.30& 	0.44 & 	9.6& \\ %\addlinespace
					&   IPCW KM    && 0.03	  (8.1) &	$0.20 $&	$0.20 $&	$4.1 $  & & 0.12  (2.4)& 	0.25& 	0.25& 	7.7 \\ %\addlinespace
					&   IPCW Cox    &&  	0.01  	(2.7)	 &$0.21$ &$	0.22 $&$	4.4$ &  & 0.07  	(1.4)& 	0.26& 	0.26& 	7.3\\
					\addlinespace \cmidrule(lr){2-12}
					
					&  Full &&  	0 	(0)&	0.17&	0.17&	2.9 & &    0  ( 0) &  0.24 &  0.24 &  5.8\\%\addlinespace
					\multirow{3}{*}{40\%}  & CC & &  0.02	  ( 5.4) & 	0.23&  0.24&  	5.3  &&   0.03   (0.6)	 &  0.32 &  0.32 &  10  \\%\addlinespace 
					&  IPCW  && 0.05 	(13.5)&  	0.22&  	0.24&  	5.1 &&    0.40  	(2.8) &  	0.28 &  	0.31 &  	24\\ %\addlinespace
					&   IPCW KM    & &  0.09   	(24) &  0.20&  	0.26&  	4.8 &&  0.20   	(4) &  	0.27 &  	0.30 &  	11\\ %\addlinespace
					&   IPCW Cox   &&  0.05	  (13.5)	&  0.21&  	0.27&  	4.6  &&     0.20    	(5) &  	0.26 &  	0.31 &  	1.0
					\\\addlinespace \cmidrule(lr){2-12}
					
					&  Full && 	0 	(0) &	$0.17$ &	$0.17$ &	$2.9$ &&  0	  (0)	 &0.24 & 0.24 &	5.8\\%\addlinespace
					\multirow{3}{*}{65\%}  & CC &&  0.02	(5.4)	& $0.30$	 &$0.30 $&	$9.0 $ & & 0.03	(0.6) &	0.43 &	0.44 &	19 \\%\addlinespace 
					&  IPCW  &&  0.05  	(13.5) & $0.24$ &	$0.33$ &$	6.0$  &&  0.20 (4) &	0.30 &	0.51 &	13	\\ %\addlinespace
					&   IPCW KM   & & 0.11	 (30) &	$0.20 $&	$0.32 $ &	$5.2 $& &  0.30 	(6) &	0.27 &	0.41 &	16\\ %\addlinespace
					&   IPCW Cox   & &  	0.08 	(21)	 &$0.21$ &$	0.33 $ & $	5.1$& &  0.20 	(4) &	0.26 &	0.37 &	11\\
					\bottomrule
				\end{tabular} 
				\begin{tablenotes}
					\footnotesize
					\item *{\bf Note}: Bias, SD, and SE provided in $10^{-2}$; MSE provided in $10^{-6}$. 
					\item  CC: complete-case analysis; IPCW: inverse probability of censoring weights; 
					 IPCW KM: IPCW via Kaplan Meier estimator; IPCW Cox: IPCW via a Cox proportional hazards model.
				\end{tablenotes}
			\end{threeparttable}

		\end{center} 
	\end{table} 
			\end{landscape}	
		%%%%%%%%%%%%%%%%%%%%%%%%%%%%%%%%%%%%%%%%%%%% %%%%%%%%%%%%%%%%%%%%%%
		%%%%%%%%%%%%%%%%%%%%%%
%		\subsection{\kern-1em:~~Another subsection} 
		
		%%%%%%%%%%%%%%%%%%%%%%%%%%%%%%%%%%%%%%%%%%%%
		%%%%%%%%%%%%%%%%%%%%%%%%%%%%%%%%%%%%%%%%%%%%
%		\section{\kern-1em:~~Appendix 2}\label{sec:app2} 
%		\end{appendices}

	%\vspace{0.9cm}
\end{document}